\documentclass[letterpaper,english,compsoc,10pt,final,conference,letterpaper]{IEEEtran}
\usepackage[T1]{fontenc}
\usepackage[utf8]{inputenc}
\usepackage{color}
\usepackage{float}
\usepackage{url}
\usepackage{amsmath}
\usepackage{amssymb}
\usepackage{graphicx}

\makeatletter

\floatstyle{ruled}
\newfloat{algorithm}{tbp}{loa}
\providecommand{\algorithmname}{Algorithm}
\floatname{algorithm}{\protect\algorithmname}

\usepackage{url}
\usepackage{flushend}
\usepackage{algorithmic}

\definecolor{green}{RGB}{217, 83, 25}

\usepackage{textpos}

\makeatother

\usepackage{babel}
\begin{document}
\title{Parallel-in-Time Kalman Smoothing Using Orthogonal Transformations}
\author{\IEEEauthorblockN{Shahaf Gargir and Sivan Toledo}
\IEEEauthorblockA{Blavatnik School of Computer Science and AI\\Tel Aviv University,  Israel}}

\maketitle
\global\long\def\exact#1{#1}%

\global\long\def\estimate#1{\hat{#1}}%

\global\long\def\weighted#1{\check{#1}}%

\global\long\def\cov{\operatorname{cov}}%

\global\long\def\expectation{\operatorname{E}}%

\begin{abstract}
We present a numerically-stable parallel-in-time linear Kalman smoother.
The smoother uses a novel highly-parallel QR factorization for a class
of structured sparse matrices for state estimation, and an adaptation
of the SelInv selective-inversion algorithm to evaluate the covariance
matrices of estimated states. Our implementation of the new algorithm,
using the Threading Building Blocks (TBB) library, scales well on
both Intel and ARM multi-core servers, achieving speedups of up to
47x on 64 cores. The algorithm performs more arithmetic than sequential
smoothers; consequently it is 1.8x to 2.5x slower on a single core.
The new algorithm is faster and scales better than the parallel Kalman
smoother proposed by Särkkä and García-Fernández in 2021.

\end{abstract}

\section{Introduction}

\begin{textblock*}{\textwidth}(0cm, -12.5cm) 
    \footnotesize\tt
    \noindent\textcopyright~2025 IEEE. Personal use of this material is permitted. 
    Permission from IEEE must be obtained for all other uses, in any current or future media, 
    including reprinting/republishing this material for advertising or promotional purposes, 
    creating new collective works, for resale or redistribution to servers or lists, 
    or reuse of any copyrighted component of this work in other works.
    \\
    Accepted to the IEEE International Parallel \& Distributed Processing
    Symposium (IPDPS), 2025.
\end{textblock*}Kalman smoothing and filtering aim to estimate a sequence of (usually
hidden) states of a dynamic system from noisy and often indirect observations
of the different states. The dynamic system is defined by inexact
evolution or state equations that relate each state to its immediate
predecessor. Noisy (inexact) observation equations relate each vector
of simultaneous observations to the state at that time. Both the evolution/state
equations and the observation equations can be linear or nonlinear\footnote{In this article, the term \emph{linear} is used to describe the evolution
and observation equations, not running times of algorithms.}. Kalman filters and smoothers estimate the states using generalized
least-squares minimization of the noise (error) terms in the equations.
The weighting of the noise terms is determined by their assumed variance/covariance.
The minimization is exact when the equations are linear; it is often
inexact when some of the equations are nonlinear.

In Kalman smoothing, we process an entire batch of observations that
has been collected. Thus, the estimation of each particular state
depends on observations of that state and all its predecessors in
the batch, as well as on observations of all successor states. Kalman
filtering is a simpler variant in which each state is estimated by
observations of itself and of predecessors. Kalman filtering has numerous
applications in which real-time state estimates are required. Kalman
smoothing is used to post process data to obtain the best possible
estimates of whole trajectories.

The first efficient linear Kalman filtering algorithm was invented
in 1960~\cite{Kalman:1960:KalmanFilter}. It was extended to smoothing
a few years later~\cite{10.2514/3.3166}. Until recently, all Kalman
filtering and smoothing algorithms have been highly sequential, processing
one state at a time. Filtering algorithms use a single forward-in-time
pass over the data (the observations). Sequential linear smoothing
algorithms consist of one forward pass and one backward pass. In 2021,
Särkkä and García-Fernández introduced the first parallel-in-time
linear Kalman smoother~\cite{10.1109/TAC.2020.2976316}. It is based
on a clever restructuring of both the forward and backward passes
in a conventional Kalman smoother as generalized prefix sums with
appropriately-defined associative operations, enabling the use of
a parallel-prefix algorithm to estimate the states and their covariance
matrices. 

In this paper we introduce a novel and completely different parallel
linear Kalman smoother with key advantages over the Särkkä and García-Fernández
smoother, including higher performance in all cases. Our algorithm,
which we refer to as \emph{Odd-Even} Kalman smoothing, is based on
a specialized sparse QR factorization of the coefficient matrix of
the linear least-squares problem that underlies Kalman smoothing.
The factorization is highly parallel thanks to a recursive odd-even
permutation of block columns of the matrix. This idea is inspired
by odd-even reduction (also called cyclic reduction), a family of
algorithms for solving block tridiagonal systems of linear equations~\cite{10.1137/0707049,10.1137/0713042}.
While this QR factorization allows us to estimate the smoothed states
efficiently in parallel, it does not provide a way to compute the
covariance matrices of these estimates.

To compute the covariance matrices of the estimates, we adapt an algorithm
called SelInv~\cite{10.1145/1916461.1916464}. This algorithm uses
the sparse triangular factor of a symmetric matrix to computes certain
blocks of the inverse of the matrix. We show how to adapt this algorithm
to the Kalman smoothing case, that it computes the necessary covariance
matrices, and that in our case it is both efficient and highly parallel.

We implemented two variants of this algorithm, as well as the Särkkä
and García-Fernández algorithm, in a state-of-the-art parallel-programming
environment. We tested them on multi-core servers with 36, 56, and
64 cores. (This is also the first detailed speedup report on the Särkkä
and García-Fernández algorithm; see~\cite{doi:10.1109/ICASSP39728.2021.9413364}
for a report on its performance on GPUs.) We also implemented the
best sequential variants of the algorithms. We show that the parallel
algorithms outperform the sequential ones and that our new algorithm
is almost always faster than the Särkkä and García-Fernández algorithm.
On the other hand, the experiments show that the parallel algorithms
do have a constant work overhead, in the sense that they perform more
arithmetic than the sequential ones, between 1.8x and 2.5x for the
two variants of our algorithm and 1.8--2.6x for the algorithm of
Särkkä and García-Fernández.

The rest of this article is structured as follows. Section~\ref{sec:Background}
describes background and related work. Section~\ref{sec:Parallel-QR}
describes our new parallel QR factorization. Section~\ref{sec:Computing-the-Covariance}
explains how we compute the covariance matrices of the estimated states.
We describe our implementation and experimental results in Section~\ref{sec:Implementation-and-Experimentation},
and we present our conclusions in Section~\ref{sec:Conclusions}.

\section{\label{sec:Background}Background and Related Work}

\subsection{Kalman Filtering and Smoothing Problems}

Kalman smoothing estimates all the states of a discrete-in-time dynamic
system that has been observed for some time~\cite{BayesianFilteringAndSmoothing2ndEd2023}.
We denote the instantaneous state of the system at time $t_{i}$ by
$\exact u_{i}\in\mathbb{R}^{n_{i}}$. We refer to $u_{i}$ as the
\emph{state} of state $i$. The state $\exact u_{i}$ satisfies a
recurrence called the \emph{evolution equation} or \emph{state equation}
and possibly a constraint called the \emph{observation equation}.
We do not require all the states to have the same dimension, although
the uniform-dimension case is very common. The evolution equation
has the form

\begin{equation}
H_{i}\exact u_{i}=\mathcal{F}_{i}\left(\exact u_{i-1}\right)+c_{i}+\epsilon_{i}\;,\label{eq:kalman-evolution-linear}
\end{equation}
where $H_{i}\in\mathbb{R}^{\ell_{i}\times n_{i}}$ is a known full-rank
matrix, $\mathcal{F}_{i}:\mathbb{R}^{n_{i-1}}\rightarrow\mathbb{R}^{\ell_{i}}$
is a known function (often assumed to be continuously differentiable),
$c_{i}\in\mathbb{R}^{\ell_{i}}$ is a known vector that represents
external forces, and $\epsilon_{i}$ is an unknown noise or error
vector. The matrix $H_{i}$ is often assumed to be the identity matrix,
but we do not require this (and do not require it to be square). A
rectangular $H_{i}$ allow modeling systems in which the dimension
of the state vector increases or decreases~\cite{10.1145/3699958}.
Obviously, the first state $u_{0}$ that we model is not defined by
an evolution recurrence.

Some of the state vectors $u_{i}$ (but perhaps not all) also satisfy
an \emph{observation} \emph{equation} of the form
\begin{equation}
o_{i}=\mathcal{G}_{i}\left(\exact u_{i}\right)+\delta_{i}\;,\label{eq:kalman-observation-linear}
\end{equation}
where $\mathcal{G}_{i}:\mathbb{R}^{n_{i}}\rightarrow\mathbb{R}^{m_{i}}$
is a known function (again usually assumed to be continuously differentiable),
$o_{i}\in\mathbb{R}^{m_{i}}$ is a known vector of observations (measurements),
and $\delta_{i}$ represents unknown measurement errors or noise.
The dimension $m_{i}$ of the observation of $u_{i}$ can vary; it
can be smaller than $n_{i}$ (including zero), equal to $n_{i}$,
or greater than $n_{i}$. 

We use $u$, $b$, and $e$ to denote the concatenations 
\[
u=\left[\begin{array}{c}
\exact u_{0}\\
\exact u_{1}\\
\exact u_{2}\\
\vdots\\
\exact u_{k-1}\\
\exact u_{k}
\end{array}\right]\;,\quad b=\left[\begin{array}{c}
o_{0}\\
c_{1}\\
o_{1}\\
\vdots\\
\vdots\\
c_{k}\\
o_{k}
\end{array}\right]\;,\quad e=\left[\begin{array}{c}
\delta_{0}\\
\epsilon_{1}\\
\delta_{1}\\
\vdots\\
\vdots\\
\epsilon_{k}\\
\delta_{k}
\end{array}\right]\;.
\]

We assume that the $\epsilon_{i}$ terms and the $\delta_{i}$ terms
are zero-mean random vectors with known covariance matrices 
\[
K_{i}=\cov\left(\epsilon_{i}\right)=\expectation\left(\epsilon_{i}\epsilon_{i}^{T}\right)\;,\quad L_{i}=\cov\left(\delta_{i}\right)=\expectation\left(\delta_{i}\delta_{i}^{T}\right)
\]
that are otherwise uncorrelated,
\begin{eqnarray*}
\expectation\left(\epsilon_{i}\delta_{j}^{T}\right) & = & 0\;\text{for all}\;i\;\text{and}\;j\\
\expectation\left(\epsilon_{i}\epsilon_{j}^{T}\right) & = & 0\;\text{for}\;i\neq j\\
\expectation\left(\delta_{i}\delta_{j}^{T}\right) & = & 0\;\text{for}\;i\neq j\;.
\end{eqnarray*}
That is, we assume that the covariance matrix of $e$ is block diagonal,
\begin{equation}
\cov(e)=\left[\begin{array}{cccccc}
L_{0}\\
 & K_{1}\\
 &  & L_{1}\\
 &  &  & \ddots\\
 &  &  &  & K_{k}\\
 &  &  &  &  & L_{k}
\end{array}\right]\;.\label{eq:structure-of-cov-e}
\end{equation}

We denote the inverse factors of $L_{i}$ and $K_{i}$ by $V_{i}^{T}V_{i}=K_{i}^{-1}$
and $W_{i}^{T}W_{i}=L_{i}^{-1}$. If we also assume that $e$ is Gaussian,
then
\begin{multline}
\begin{bmatrix}\estimate{\exact u}_{0}\\
\vdots\\
\estimate{\exact u}_{k}
\end{bmatrix}=\arg\min\left(\sum_{i=1}^{k}\left(V_{i}\left(H_{i}\exact u_{i}-\mathcal{F}_{i}\left(\exact u_{i-1}\right)-c_{i}\right)\right)^{2}\right.\\
\left.+\sum_{i=0}^{k}\left(W_{i}\left(o_{i}-\mathcal{G}_{i}\left(\exact u_{i}\right)\right)\right)^{2}\right)\label{eq:global-nonlinear-ls}
\end{multline}
is the maximum-likelihood estimator of $u$. If $e$ is not necessarily
Gaussian but all $\mathcal{F}_{i}$ and $\mathcal{G}_{i}$ are linear
(matrices), the estimator~(\ref{eq:global-nonlinear-ls}) is the
minimum-variance unbiased estimator~\cite{KailathSayedHassibiLinearEstimation}.
We normally need not only the $\estimate u_{i}$s, but also their
covariance matrices $\cov(\estimate u_{i})$. We denote by ${\cal A}(u)$
the nonlinear function that returns the vector 
\begin{equation}
{\cal A}(u)=\left[\begin{array}{c}
\mathcal{G}_{0}(\exact u_{0})\\
H_{1}u_{1}-\mathcal{F}_{1}(u_{0})\\
\mathcal{G}_{1}(\exact u_{1})\\
H_{2}u_{2}-\mathcal{F}_{2}(u_{1})\\
\vdots\\
\vdots\\
H_{k}u_{k}-\mathcal{F}_{k}(u_{k-1})\\
\mathcal{G}_{k}(\exact u_{k})
\end{array}\right]\;.\label{eq:global-nonlinear-function}
\end{equation}
With this notation, the least-squares estimator is
\begin{equation}
\estimate u=\arg\min_{u}\left\Vert U\left({\cal A}(u)-b\right)\right\Vert _{2}^{2}\label{eq:nonlinear-generalized-least-squares}
\end{equation}
where $U^{T}U=\cov(e)^{-1}$.

In linear Kalman filtering and smoothing, the functions $\mathcal{F}_{i}$
and $\mathcal{G}_{i}$ are real matrices.

\subsection{Kalman Filtering/Smoothing Paradigms}

The original Kalman filter~\cite{Kalman:1960:KalmanFilter} from
1960 for linear dynamic systems and many of its variants can be derived
as formulas that track the expectation and covariance of the state.
These algorithms can be extended to smoothers, sometimes called RTS
smoothers, using a backward (in time) sweep, to propagate future information
back~\cite{10.2514/3.3166}. Good modern references for these algorithms
include~\cite{10.1137/100799666,BayesianFilteringAndSmoothing2ndEd2023,KailathSayedHassibiLinearEstimation}.
Another family of Kalman filter algorithms, sometimes called \emph{information
filters}, track the expectation and the inverse of the covariance
matrices of the states. Some variants of these algorithms track a
Cholesky factor of the covariance matrix or its inverse. In general,
all of these algorithms require the expectation and covariance of
the initial state to be known, and some of them can easily handle
singular $K_{i}$s and/or singular $L_{i}$s. Most cannot handle rectangular
$H_{i}$s.

In 1972 Duncan and Horn realized that when $\cov(e)$ is nonsingular,
the Kalman-smoothed trajectory of a linear dynamic system can be expressed
as a generalized linear least-squares problem~(\ref{eq:nonlinear-generalized-least-squares})
with ${\cal A}$ replaced by a matrix $A$~\cite{DucanHornKalman}.
In 1977 Paige and Saunders developed a Kalman filter and smoother
based on a specialized $UA=QR$ factorization for the special sparsity
structure of $UA$~\cite{PaigeSaunders:1977:Kalman}. They also introduced
an algorithm to compute $\cov(\estimate u_{i})$ using a sequence
of orthogonal transformations of the $R$ factor. Their algorithm
requires $\cov(e)$ to be nonsingular, but it does not require that
the expectation of the initial state be known. (Their algorithm also
assumes that $H_{i}=I$, but can be easily extended to arbitrary full-rank
$H_{i}$s~\cite{10.1145/3699958}.) Due to the use of orthogonal
transformation, this algorithm is likely to be more numerically stable
than other Kalman filters and smoothers.

All of these algorithms have the same asymptotic efficiency. Assuming
that $n_{i}=\Theta(n)$, $\ell_{i}=\Theta(n)$, and $m_{i}=O(n)$
for some state-dimension $n$, Kalman filtering requires $\Theta(n^{3})$
operations per step for a total of $\Theta(kn^{3})$ operations and
$\Theta(n^{2})$ storage, and Kalman smoothing requires $\Theta(kn^{3})$
operations and $\Theta(kn^{2})$ storage\footnote{We use big-$O$ to denote asymptotic upper bounds and big-$\Theta$
to denote asymptotic proportionality; see~\cite[Chapter~3]{CLRS2nd}.}.

When the $\mathcal{F}_{i}$s and/or $\mathcal{G}_{i}$s are nonlinear,
the minimum $\estimate u$ can be determined using an iterative Gauss-Newton-type
algorithm~\cite{Bjorck:1996:NML2e}. Each step of the algorithm solves
a \emph{linear} generalized least squares problems of the form~(\ref{eq:global-nonlinear-function})
but with the $\mathcal{F}_{i}$s and $\mathcal{G}_{i}$s replaced
by matrices. These matrices are the Jacobians $F_{i}=\partial\mathcal{F}_{i}/\partial u_{i}$
and $G_{i}=\partial\mathcal{G}_{i}/\partial u_{i}$ evaluated at the
estimate of $u$ produced by the previous iteration (or Jacobians
concatenated with a diagonal matrix)~\cite{10.1137/0804035,10.1109/ICASSP40776.2020.9054686}.
That is, Kalman smoothing of a nonlinear dynamic system can be algorithmically
reduced to Kalman smoothing of a sequence of linear dynamic systems.
The covariance of the states of these linear problems is not needed.
(Gauss-Newton algorithms also require an initial guess for $u$, which
in the case of dynamic systems can be obtained using an extended or
unscented Kalman filter~\cite{BayesianFilteringAndSmoothing2ndEd2023}
or one of their variants; this step is beyond the scope of this paper.)

\subsection{Parallel-in-Time Kalman Smoothing}

Kalman filters are incremental (streaming) and therefore inherently
sequential, in the sense that they handle step after step. Until recently,
all Kalman smoothing algorithms were also sequential, handling step
after step with at least one forward sweep in time and one backward
sweep in time.

In both filtering and smoothing, the $\Theta(n^{3})$ operations in
each step can be parallelized~\cite{10.1109/TCST.2011.2176735},
although typical values of $n$ are not large enough to merit that.
Generally speaking, these $\Theta(n^{3})$ operations consist of matrix
multiplications, matrix factorizations, and solutions of triangular
linear systems of equations with multiple right-hand sides. The length
of the critical path in practical parallel factorization algorithms
and triangular solvers is $\Theta(n)$ (the term \emph{critical path}
refers to the longest chain of dependences; it is also known as \emph{span}
or \emph{depth} of the computation).

In 2021, Särkkä and García-Fernández introduced the first parallel-in-time
linear Kalman smoother~\cite{10.1109/TAC.2020.2976316}. Their algorithm
is based on a restructuring of the forward and backward sweeps of
the conventional Kalman (RTS) smoother~\cite{10.2514/3.3166} as
prefix sums of associative operations operating on the state.  Expressing
the smoothed trajectory and the covariance matrices $\cov(\estimate u_{i})$
as quantities computed by two prefix sum operations allows them to
be computed using a parallel prefix-sum (parallel scan) algorithm.
 This algorithm, as well as a variant in which covariance matrices
are represented differently, have been implemented on GPUs by Yaghoobi,
Corenflos, Hassan, and Särkkä, and have shown to be faster than the
baseline sequential algorithms~\cite{doi:10.1109/ICASSP39728.2021.9413364,doi:10.48550/arXiv.2207.00426}.

\section{\label{sec:Parallel-QR}A Parallel QR Factorization for Kalman Matrices}

We now describe our new parallel QR factorization, which forms the
basis for our new smoother. The algorithm is closely related to the
Paige-Saunders factorization, but it reorders block columns in order
to introduce parallelism.

To efficiently find the minimizer~(\ref{eq:global-nonlinear-ls}),
we need to compute, in parallel, a $QR=UAP$ factorization of a column
permutation $P$ of a block matrix $UA$ of the form

\[
UA=\small\setlength{\arraycolsep}{3.5pt}\left[\begin{array}{cccccc}
C_{0}\\
-B_{1} & D_{1}\\
 & C_{1}\\
 & -B_{2} & D_{2}\\
 &  & \ddots & \ddots\\
 &  &  & \ddots & \ddots\\
 &  &  &  & -B_{k} & D_{k}\\
 &  &  &  &  & C_{k}
\end{array}\right]
\]
where $C_{i}=W_{i}G_{i}$, $B_{i}=V_{i}F_{i}$, and $D_{i}=V_{i}H_{i}$.
The dimensions of block rows and block columns can vary. To simplify
complexity analyses, we assume that the dimensions of all the blocks
are $O(n)$ for some $n$. We allow $Q$ to be represented as a product
of orthonormal matrices.

We number block columns from $0$ to $k$.

We base the algorithm on the block odd-even reduction (or cyclic reduction)
algorithm for block tridiagonal matrices~\cite{10.1137/0707049,10.1137/0713042}.
The algorithm is recursive and it is based on a recursive reordering
of the block columns. At the top level, we order the even block columns
first, 
\[
\small\setlength{\arraycolsep}{3.5pt}\left[\begin{array}{ccccc|cccc}
C_{0} &  &  &  & \\
-B_{1} &  &  &  &  & D_{1}\\
 &  &  &  &  & C_{1}\\
 & D_{2} &  &  &  & -B_{2}\\
 & C_{2} &  &  & \\
 & -B_{3} &  &  &  &  & D_{3}\\
 &  &  &  &  &  & C_{3}\\
 &  & D_{4} &  &  &  & -B_{4}\\
 &  & \vdots &  &  &  &  & \ddots\\
 &  &  &  &  &  &  &  & D_{k-1}\\
 &  &  & \ddots &  &  &  &  & C_{k-1}\\
 &  &  &  & D_{k} &  &  &  & -B_{k}\\
 &  &  &  & C_{k}
\end{array}\right]
\]
(if $k$ is even) or, if $k$ is odd,
\[
\small\setlength{\arraycolsep}{3.5pt}\left[\begin{array}{ccccc|cccc}
C_{0} &  &  &  & \\
-B_{1} &  &  &  &  & D_{1}\\
 &  &  &  &  & C_{1}\\
 & D_{2} &  &  &  & -B_{2}\\
 & C_{2} &  &  & \\
 & -B_{3} &  &  &  &  & D_{3}\\
 &  &  &  &  &  & \vdots\\
 &  & D_{4} &  & \\
 &  & \vdots &  & \\
 &  &  & \ddots & \\
 &  &  &  & D_{k-1}\\
 &  &  &  & C_{k-1} &  &  & \ddots\\
 &  &  &  & -B_{k} &  &  &  & D_{k}\\
 &  &  &  &  &  &  &  & C_{k}
\end{array}\right]\;.
\]
We continue without loss of generality with the first case.

We compute QR factorizations of the last two nonzero block rows in
each block column in the left side and we apply the Q factors to the
two block rows. This causes one block of fill in the right part of
the matrix. That is, let 
\[
\left[\begin{array}{c}
C_{i}\\
-B_{i+1}
\end{array}\right]=Q_{i}\left[\begin{array}{c}
\tilde{R}_{i}\\
0
\end{array}\right]
\]
and so on. Our matrix is reduced to 
\[
\small\setlength{\arraycolsep}{3.5pt}\left[\begin{array}{ccccc|cccc}
\tilde{R}_{0} &  &  &  &  & X_{0}\\
0 &  &  &  &  & \tilde{D}_{1}\\
 &  &  &  &  & C_{1}\\
 & D_{2} &  &  &  & -B_{2}\\
 & \tilde{R}_{2} &  &  &  &  & X_{2}\\
 & 0 &  &  &  &  & \tilde{D}_{3}\\
 &  &  &  &  &  & C_{3}\\
 &  & D_{4} &  &  &  & -B_{4}\\
 &  & \vdots &  &  &  &  & \ddots\\
 &  &  &  &  &  &  &  & X_{k-2}\\
 &  &  &  &  &  &  &  & \tilde{D}_{k-1}\\
 &  &  & \ddots &  &  &  &  & C_{k-1}\\
 &  &  &  & D_{k} &  &  &  & -B_{k}\\
 &  &  &  & \tilde{R}_{k}
\end{array}\right]
\]
We denoted the blocks that filled by $X_{i}$. We now compute QR factorizations
of the nonzero block rows in each block column in the left side of
the matrix. Note that in the first block column, there is nothing
to do. We reach the following form:
\[
\small\setlength{\arraycolsep}{3.5pt}\left[\begin{array}{ccccc|cccc}
R_{0} &  &  &  &  & X_{0}\\
0 &  &  &  &  & \tilde{D}_{1}\\
 &  &  &  &  & C_{1}\\
 & R_{2} &  &  &  & -\tilde{B}_{2} & Y_{2}\\
 & 0 &  &  &  & Z_{2} & \tilde{X}_{2}\\
 & 0 &  &  &  &  & \tilde{D}_{3}\\
 &  &  &  &  &  & C_{3}\\
 &  & R_{4} &  &  &  & -\tilde{B}_{4}\\
 &  & \vdots &  &  &  &  & \ddots\\
 &  &  &  &  &  &  &  & X_{k-1}\\
 &  &  &  &  &  &  &  & \tilde{D}_{k-1}\\
 &  &  & \ddots &  &  &  &  & C_{k-1}\\
 &  &  &  & R_{k} &  &  &  & -\tilde{B}_{k}\\
 &  &  &  & 0 &  &  &  & Z_{k}
\end{array}\right]\;.
\]
We now permute (at least conceptually) the block rows $R_{i}$ to
the top, 
\[
\small\setlength{\arraycolsep}{3.5pt}\left[\begin{array}{ccccc|cccc}
{\color{green}R_{0}} & {\color{green}} & {\color{green}} & {\color{green}} & {\color{green}} & {\color{green}X_{0}} & {\color{green}} & {\color{green}} & {\color{green}}\\
{\color{green}} & {\color{green}R_{2}} & {\color{green}} & {\color{green}} & {\color{green}} & {\color{green}-\tilde{B}_{2}} & {\color{green}Y_{2}} & {\color{green}} & {\color{green}}\\
{\color{green}} & {\color{green}} & {\color{green}R_{4}} & {\color{green}} & {\color{green}} & {\color{green}} & {\color{green}-\tilde{B}_{4}} & {\color{green}} & {\color{green}}\\
{\color{green}} & {\color{green}} & {\color{green}} & {\color{green}\ddots} & {\color{green}} & {\color{green}} & {\color{green}} & {\color{green}} & {\color{green}}\\
{\color{green}} & {\color{green}} & {\color{green}} & {\color{green}} & {\color{green}R_{k}} & {\color{green}} & {\color{green}} & {\color{green}} & {\color{green}-\tilde{B}_{k}}\\
 &  &  &  &  & \tilde{D}_{1}\\
 &  &  &  &  & C_{1}\\
 &  &  &  &  & Z_{2} & \tilde{X}_{2}\\
 &  &  &  &  &  & \tilde{D}_{3}\\
 &  &  &  &  &  & C_{3}\\
 &  &  &  &  &  &  & \ddots\\
 &  &  &  &  &  &  &  & X_{k-1}\\
 &  &  &  &  &  &  &  & \tilde{D}_{k-1}\\
 &  &  &  &  &  &  &  & C_{k-1}\\
 &  &  &  &  &  &  &  & Z_{k}
\end{array}\right]\;.
\]
The nonzero blocks in the first $k/2$ block rows, colored red, now
permanent blocks of the overall R factor. 

We can almost recurse. The bottom part of the right side is almost
identical to the structure that we started with, except that in each
block column, there are two unique block rows, not one. We can still
argue that the number of rows in 
\[
\left[\begin{array}{c}
\tilde{D}_{1}\\
C_{1}
\end{array}\right],\;\left[\begin{array}{c}
\tilde{D}_{3}\\
C_{3}
\end{array}\right],\;\ldots
\]
is asymptotically $O(n)$, but as the recursion gets deeper, this
asymptotic claim no longer holds. To restore the invariant on the
size of these blocks, we replace them by their R factor, to reach
the following structure:
\begin{equation}
\small\setlength{\arraycolsep}{3.5pt}\left[\begin{array}{ccccc|cccc}
{\color{green}R_{0}} & {\color{green}} & {\color{green}} & {\color{green}} & {\color{green}} & {\color{green}X_{0}} & {\color{green}} & {\color{green}} & {\color{green}}\\
{\color{green}} & {\color{green}R_{2}} & {\color{green}} & {\color{green}} & {\color{green}} & {\color{green}-\tilde{B}_{2}} & {\color{green}Y_{2}} & {\color{green}} & {\color{green}}\\
{\color{green}} & {\color{green}} & {\color{green}R_{4}} & {\color{green}} & {\color{green}} & {\color{green}} & {\color{green}-\tilde{B}_{4}} & {\color{green}} & {\color{green}}\\
{\color{green}} & {\color{green}} & {\color{green}} & {\color{green}\ddots} & {\color{green}} & {\color{green}} & {\color{green}} & {\color{green}} & {\color{green}}\\
{\color{green}} & {\color{green}} & {\color{green}} & {\color{green}} & {\color{green}R_{k}} & {\color{green}} & {\color{green}} & {\color{green}} & {\color{green}-\tilde{B}_{k}}\\
 &  &  &  &  & \tilde{C}_{1}\\
 &  &  &  &  & 0\\
 &  &  &  &  & Z_{2} & \tilde{X}_{2}\\
 &  &  &  &  &  & \tilde{C}_{3}\\
 &  &  &  &  &  & 0\\
 &  &  &  &  &  &  & \ddots\\
 &  &  &  &  &  &  &  & X_{k-1}\\
 &  &  &  &  &  &  &  & \tilde{C}_{k-1}\\
 &  &  &  &  &  &  &  & 0\\
 &  &  &  &  &  &  &  & Z_{k}
\end{array}\right]\;.\label{eq:half-factored}
\end{equation}
The zero block rows are permuted to the end. We now have the same
block structure that we started with and we recurse.

Let us analyze the parallelism in this process. Let $i$ be an even
index. All the 2-block $QR$ factorizations of the 
\[
\left[\begin{array}{c}
C_{i}\\
-B_{i+1}
\end{array}\right]
\]
can be computed in parallel and the $Q$ factors can be applied to
the even columns in parallel as these factorizations operate on disjoint
pairs of block rows. The $QR$ factorizations of the submatrices
\[
\left[\begin{array}{c}
D_{i}\\
\tilde{R}_{i}
\end{array}\right]
\]
can also be computed and applied concurrently, and in parallel we
can compute and apply the $QR$ factorizations of the submatrices
\[
\left[\begin{array}{c}
\tilde{D}_{i-1}\\
C_{i-1}
\end{array}\right]\;.
\]
Therefore, computing the even block columns of $R$ requires two steps,
each involving a number of concurrent $QR$ factorizations of $2$-by-$1$
block matrices.

\begin{figure}
\hfill{}\includegraphics[width=0.8\columnwidth]{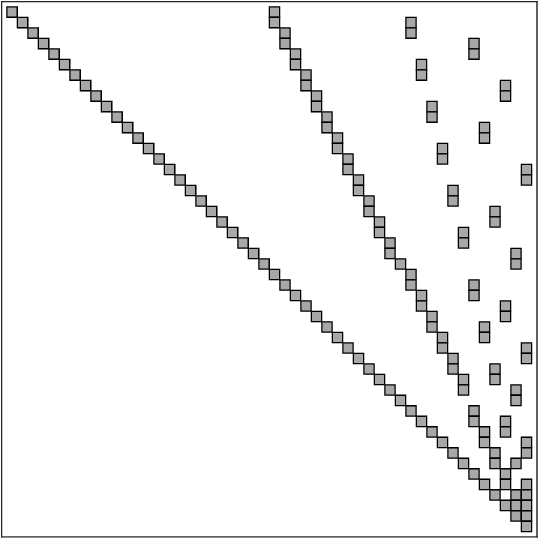}\hfill{}

\caption{\label{fig:R-structures}The structure of $R$ in the odd-even algorithm.
The problem consisted of $k=50$ states. Each gray square represents
an $n$-by-$n$ nonzero block. }
 
\end{figure}

Figure~\ref{fig:R-structures} shows the nonzero block structure
of the resulting upper-triangular factor $R$.

\subsection{Transformation of Right-Hand Sides and Back Substitution}

We have shown how to compute a $QR=UAP$ factorization of a column
permutation $P$ of $UA$ to minimize $\|U(Au-b)\|_{2}$. Orthogonal
transformations and block-row permutations are also applied to $Ub$,
to form a transformed right-hand side that we denote $Q^{T}Ub$. This
is essentially trivial since the algorithm applies in parallel batches
of orthogonal transformations that each modifies two block rows, and
the pairs are always disjoint.

Once we complete the factorization and the transformation of $Ub$,
we can solve for the optimal $u$, $RP^{T}\hat{u}=Q^{T}Ub$. We use
the same recursive structure to solve for $y=P^{T}\hat{u}$ and then
permute to obtain $\hat{u}$. Assume that we already determined recursively
the right (bottom) half of $y$. We multiply each computed block of
$y$ by one or two corresponding blocks in the upper-right part of~(\ref{eq:half-factored}),
sum up to two such sub-vectors, subtract from the corresponding blocks
of $Q^{T}Ub$, and solve the triangular block linear system that generates
each block of $y$.

\subsection{\label{subsec:Data-Structures}Data Structures}

We store the inputs to the algorithm, intermediate matrices and vectors,
and the outputs in an array of structures. Each structure represents
one step and the matrix and vector blocks associated with it. We initialize
the data structure with the inputs: $F_{i}$, $H_{i}$, $G_{i}$,
$K_{i}$, $L_{i}$, $c_{i}$, and $o_{i}$. The use of an array of
structure allows loading the inputs or creating them in parallel.
In particular, in a Gauss-Newton-type nonlinear solver, the inputs
are the Jacobians $F_{i}=\partial\mathcal{F}_{i}/\partial u_{i}$
and $G_{i}=\partial\mathcal{G}_{i}/\partial u_{i}$ evaluated at the
estimate of $u$ produced by the previous iteration (or Jacobians
concatenated with a diagonal matrix)~\cite{10.1137/0804035,10.1109/ICASSP40776.2020.9054686},
and the residual vector; both can be trivially evaluated in parallel.

Our implementation of the Särkkä and García-Fernández also includes
a concurrent-set data structure. We use it to ensure that all memory
allocated in the scope of parallel scan operations is released when
they complete. 

\subsection{Analysis of Work and Critical Path (Depth)}

We assume that the QR factorization of a matrix with $m$ rows and
$n$ columns requires $\Theta(mn^{2})$ arithmetic operations and
memory accesses and that the critical path of the factorization has
length $\Theta(n\log n)$ (this quantity is also known as span or
depth) . We assume that the number of rows and columns in $F_{i}$
is $\Theta(n)$ and that the number of rows in $G_{i}$ is $O(n)$
for some $n$, the typical dimension of state vectors.

Our algorithm starts with a column-reordering phase. We implement
this phase by permuting the block-column index array, which is initialized
to $0,\ldots,k$. The permutation performs $\Theta(k)$ work and can
be performed in perfect parallelism using an auxiliary array. 

We now analyze the work and critical path of the algorithm. The factorization
\[
Q_{i}^{T}\left[\begin{array}{c}
C_{i}\\
-B_{i+1}
\end{array}\right]=\left[\begin{array}{c}
\tilde{R}_{i}\\
0
\end{array}\right]
\]
requires $\Theta(n^{3})$ work and has critical path $\Theta(n\log n)$.
The application of $Q_{i}^{T}$ to a block column in the right side
of matrix 
\[
Q_{i}^{T}\left[\begin{array}{c}
0\\
D_{i+1}
\end{array}\right]=\left[\begin{array}{c}
X_{i}\\
\tilde{D}_{i+1}
\end{array}\right]
\]
incurs similar costs, $\Theta(n^{3})$ work and critical path $O(n\log n)$.
The critical path can probably be reduced, depending on how $Q_{i}^{T}$
is represented, but this does not modify the overall asymptotic costs
of the algorithm. 

These operations can be conducted in parallel on all the block columns,
since each of these QR factorizations modifies a pair of block rows,
and the pairs are disjoint. Therefore, the total work in this step
is $\Theta(kn^{3})$ and the total critical path is still $\Theta(n\log n)$.

The next step is to compute and then apply the factorizations
\begin{eqnarray*}
\tilde{Q}_{i}^{T}\left[\begin{array}{c}
D_{i}\\
\tilde{R}_{i}
\end{array}\right] & = & \left[\begin{array}{c}
R_{i}\\
0
\end{array}\right]\\
\tilde{Q}_{i}^{T}\left[\begin{array}{cc}
-B_{i}\\
 & X_{i}
\end{array}\right] & = & \left[\begin{array}{cc}
-\tilde{B}_{i} & Y_{i}\\
Z_{i} & \tilde{X}_{i}
\end{array}\right]\;.
\end{eqnarray*}
These operations have the same work and critical-path bounds and
all of them can be carried out concurrently.

The last step before we recurse is to concurrently (for all $i$)
factor
\[
\hat{Q}_{i}^{T}\left[\begin{array}{c}
\tilde{D}_{i+1}\\
C_{i+1}
\end{array}\right]=\left[\begin{array}{c}
\tilde{C}_{i+1}\\
0
\end{array}\right]\;,
\]
again with the same work and critical-path bounds.

This gives us the following recurrences for work and critical path,
\begin{eqnarray*}
T_{1}\left(k,n\right) & = & 3k\Theta\left(n^{3}\right)+T_{1}\left(\frac{k}{2},n\right)\\
 & = & \Theta\left(kn^{3}\right)+T_{1}\left(\frac{k}{2},n\right)\\
T_{\infty}\left(k,n\right) & = & 3\Theta\left(n\log n\right)+T_{\infty}\left(\frac{k}{2},n\right)\\
 & = & \Theta\left(n\log n\right)+T_{\infty}\left(\frac{k}{2},n\right)\;.
\end{eqnarray*}
In the base case $k=O(1)$ the work and critical-path bounds are $\Theta(n^{3})$
and $\Theta(n\log n)$. Therefore, the recurrences solve to
\begin{eqnarray*}
T_{1}\left(k,n\right) & = & \Theta\left(kn^{3}\right)\\
T_{\infty}\left(k,n\right) & = & \Theta\left(\log k\cdot n\log n\right)\;.
\end{eqnarray*}

The asymptotic work bound is the same as the work bound for the original
Paige-Saunders algorithm. The constant factors are larger, but by
a fairly small constant. We explore this issue experimentally below.
The critical path of the original Paige-Saunders algorithm is $\Theta(k\cdot n\log n)$,
dramatically worse than our new algorithm for large $k$.

\section{\label{sec:Computing-the-Covariance}Computing the Covariance Matrices
of the Estimates}

Conventional Kalman filters and smoothers track the state $\estimate u_{i}$
and its covariance $\cov(\estimate u_{i})$ simultaneously. The Paige-Saunders
QR-based smoother produces all the $\estimate u_{i}$ without computing
their covariance matrices. Paige and Saunders proposed a second algorithm
that recovers the $\cov(\estimate u_{i})$ matrices from the block-bidiagonal
$R$ factor of $QR=UA$. This algorithm is clever and elegant, but
there is no apparent way to extend it to our factorization of an odd-even
column-permuted $UA$, since the resulting $R$ factor is no longer
block bidiagonal. 

Therefore, we rely on a completely different way of computing $\cov(\estimate u_{i})$.
It is well known that when 
\begin{equation}
\estimate u=\arg\min_{u}\left\Vert U\left(Au-b\right)\right\Vert _{2}^{2}\label{eq:nonlinear-generalized-least-squares-1}
\end{equation}
and $UA=QR$ is a thin QR factorization ($R$ square), $\cov(\estimate u)=(R^{T}R)^{-1}$~\cite{PaigeSaunders:1977:Kalman}.
The matrices we seek, $\cov(\estimate u_{i})$, are the diagonal blocks
of this matrix.

We adapt an algorithm called SelInv~\cite{10.1145/1916461.1916464}
to compute these diagonal blocks. SelInv is an efficient algorithm
to compute the diagonal of $(LDL^{T})^{-1}$ where $L$ is sparse
and unit lower triangular (has $1$s on the diagonal) and $D$ is
diagonal, or to compute the diagonal blocks of $(LDL^{T})^{-1}$ when
$L$ is sparse and block unit lower triangular (its diagonal blocks
are identities) and $D$ is block diagonal (these matrices are unrelated
to $L_{i}$ and $D_{i}$ in Sections~\ref{sec:Background} and~\ref{sec:Parallel-QR}).
We use the latter, block, variant. SelInv also computes other blocks
of these inverses but we do not need them. Its efficiency depends
on the sparsity of $L$; the sparser it is, the more efficient the
algorithm. 

We cannot apply the block variant directly to $(R^{T}R)^{-1}$. Our
$R$ factor is sparse and block upper triangular, but its diagonal
blocks are not identities. Therefore, we map $R$ onto $L$ and $D$
as follows:
\begin{eqnarray*}
D_{ii} & = & R_{ii}^{T}R_{ii}\\
L_{ii} & = & I\\
L_{ij} & = & (R_{ji})^{T}R_{jj}^{-T}\;.
\end{eqnarray*}
Due to a lack of space in this extended abstract, we omit the fairly
trivial correctness proof of this mapping.  Given this mapping, the
block variant of SelInv maps into Algorithm~\ref{alg:block-selinv-for-rtr}.
The algorithm computes all the blocks of $S=(R^{T}R)^{-1}$ that are
nonzero in $R$, including the diagonal blocks. 

\begin{algorithm}
\caption{\label{alg:block-selinv-for-rtr}Block SelInv adapted to $S=(R^{T}R)^{-1}$.}

\begin{algorithmic}
\STATE $S_{k,k} \gets R_{k,k}^{-1} R_{k,k}^{-T}$
\FOR {$j \gets k-1$ down to $0$} 
  \STATE $\cal{I} \gets $ indexes of offdiagonal nonzero blocks in block row $j$ of $R$ 
  \STATE $S_{j,\cal{I}} \gets - R^{-1}_{j,j} R_{j,\cal{I}}  S_{\cal{I},\cal{I}} $
  \STATE $S_{{\cal I},j} \gets S_{j,\cal{I}}^T$
  \STATE $S_{j,j} \gets R_{j,j}^{-1} R_{j,j}^{-T} - S_{j,\cal{I}} \left( R^{-1}_{j,j} R_{j,\cal{I}} \right)^T $
\ENDFOR
\end{algorithmic}

\end{algorithm}

In the sequential case (the Paige-Saunders algorithm), we have ${\cal I}=\{j+1\}$
in every iteration, so each iteration requires two matrix multiplications
and, three triangular solve with $n$ right-hand sides, so it preserves
the asymptotic complexity of Paige and Saunders' approach. 

\begin{algorithm}
\caption{\label{alg:odd-even-block-selinv-for-rtr}Odd-even block SelInv for
$S=(R^{T}R)^{-1}$.}

\begin{algorithmic}
\IF {$R$ has only one block column $j$} 
  \STATE $S_{j,j} \gets R_{j,j}^{-1} R_{j,j}^{-T}$
  \RETURN
\ENDIF
\STATE Recurse on odd block columns (which were permuted last)
\FOR {every even block column $j$ in parallel} 
  \STATE $\cal{I} \gets $ indexes of offdiagonal nonzero blocks in block row $j$ of $R$ 
  \STATE $S_{j,\cal{I}} \gets - R^{-1}_{j,j} R_{j,\cal{I}} S_{\cal{I},\cal{I}} $
  \STATE $S_{{\cal I},j} \gets \left( S_{j,\cal{I}} \right)^T$
  \STATE $S_{j,j} \gets R_{j,j}^{-1} R_{j,j}^{-T} - S_{j,\cal{I}} \left( R^{-1}_{j,j} R_{j,\cal{I}} \right)^T $
\ENDFOR
\end{algorithmic}
\end{algorithm}

Algorithm~\ref{alg:odd-even-block-selinv-for-rtr} presents a parallel
version of this method, specialized to the odd-even $R$ factor shown
in Equation~(\ref{eq:half-factored}). We start with the odd columns,
recursively. Once the recursion ends, we have the nonzeros blocks
of $S=(R^{T}R)^{-1}$ corresponding to nonzero blocks in the odd columns
of $R$, except in the first $\left\lceil k/2\right\rceil $ block
rows. We now process all the even rows, shown in red in Equation~(\ref{eq:half-factored}),
in parallel. We can process these rows in parallel because the sets
${\cal I}$ only include column indices that have been processed recursively.
Note that in this case, $|{\cal I}|=2$ or $|{\cal I}|=1$, so the
total arithmetic cost is higher than in the sequential bidiagonal
case, but not asymptotically higher. The critical path is again $T_{\infty}\left(k,n\right)=\Theta\left(\log k\cdot n\log n\right)$.
The $\log k$ factor is the depth of the recursion, and the $n\log n$
is the assumed depth of the triangular solves.

\section{\label{sec:Implementation-and-Experimentation}Implementation and
Experimental Results}

\begin{figure*}
\hspace{0.1\columnwidth}\includegraphics[width=0.8\columnwidth]{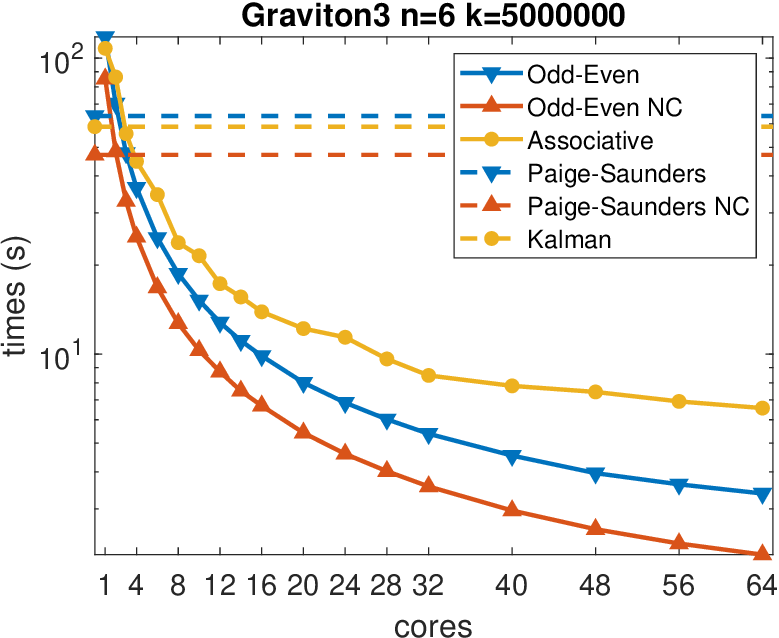}\hfill{}\includegraphics[width=0.8\columnwidth]{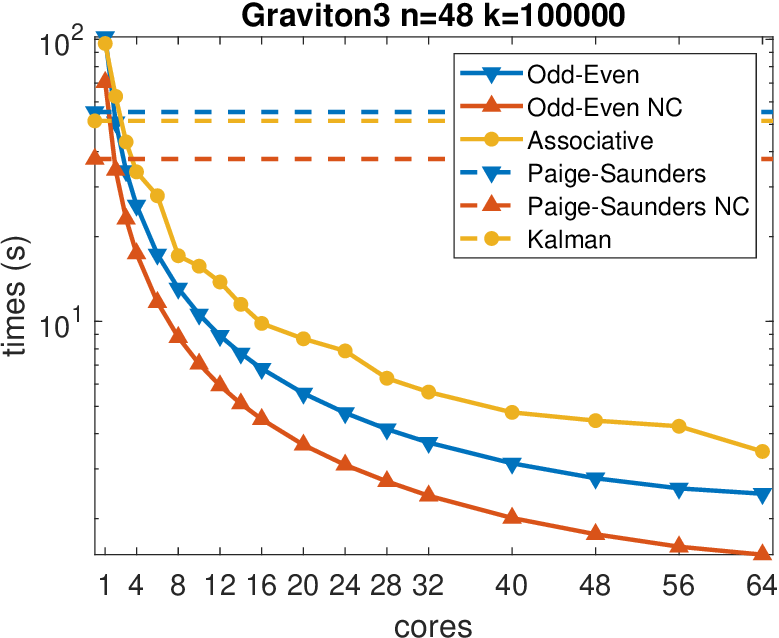}\hspace{0.1\columnwidth}\hphantom{}\\
~\\
\hphantom{}\hspace{0.1\columnwidth}\includegraphics[width=0.8\columnwidth]{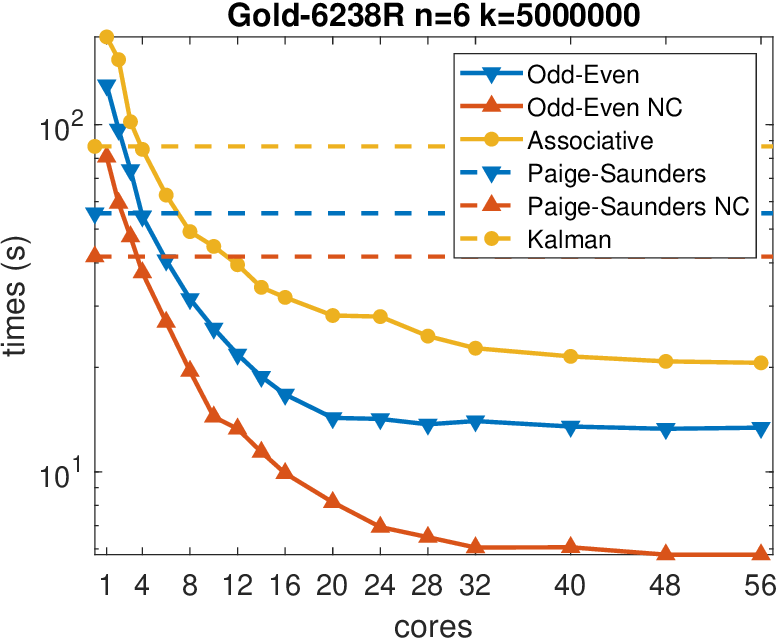}\hfill{}\includegraphics[width=0.8\columnwidth]{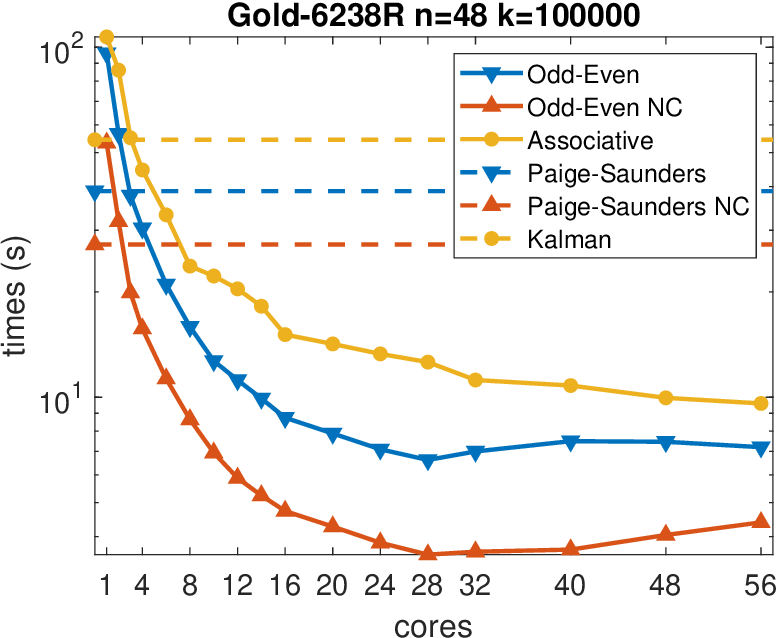}\hspace{0.1\columnwidth}

\caption{\label{fig:running-times}Running Times of all the smoothers on a
server with 64 physical cores (Graviton3) and on a server with 56
physical cores (2 Intel Xeon Gold 6238R CPUs).}
\end{figure*}

\subsection{Implementation}

We implemented all the algorithms that we described, including one
of the sequential baseline algorithms, in both Matlab and C, using
double-precision (64-bit) floating-point arithmetic. The implementation
is based on the UltimateKalman implementation of the sequential Paige-Saunders
algorithm~\cite{10.1145/3699958} and uses its API. We only report
on the performance of the C codes. We implemented the parallel algorithms
using the open-source threading building blocks (TBB) library~\cite{TBB}.
All the $\Theta(n^{3})$ matrix operations (matrix multiplications,
factorizations, etc) are implemented using calls to a high-performance
linear algebra library through the standard BLAS and LAPACK interfaces. 

TBB is a C++ library designed to enable the implementation of shared-memory
multi-core algorithms. It consists of a randomized work-stealing scheduler,
a scalable memory allocator, as well as many convenience routines~\cite{TBB}.
The scheduler is based on the Cilk scheduler and it provides similar
theoretical guarantees~\cite{10.1109/SFCS.1994.365680,10.1145/209937.209958}.
We chose to implement the algorithm with TBB for several reasons:
(1) the theoretical performance guarantees, (2) the use of a library
rather than a compiler-based parallel programming language reduces
the risk of obsolescence and improves portability, (3) TBB, like Cilk,
supports nested parallelism, which allows parallelism to be exploited
at both the top parallel-in-time level and at the linear-algebra level
below. 

Our implementations consist almost entirely of C code. We use a single
small C++ source file that contains two C-callable C++ functions.
One of these functions invokes the \texttt{tbb::parallel\_for} template
function, while the other invokes \texttt{tbb::parallel\_scan}. These
functions also instructs TBB to use a certain number of cores (operating
system threads) and to use a particular block size, the number of
iterations or data items that are performed sequentially to reduce
scheduling overheads. We use a block size of 10 unless noted otherwise.

Since it is impossible to restrict TBB to use only a single core (the
restriction mechanism works for 2 cores and up), we also compile a
sequential version of each parallel algorithm. This version replaces
the calls to \texttt{tbb::parallel\_for} and \texttt{tbb::parallel\_scan}
with simple C loops that perform the same computation sequentially.
These versions also skip other parallel overheads, in particular the
concurrent set operations that we use to ensure that all the memory
allocated within \texttt{tbb::parallel\_scan} is released.

Our test programs use TBB's scalable memory allocator, not the memory
allocator of the C library. This is done by linking them with a proxy
library, \texttt{libtbbmalloc\_proxy} that implements all the C and
C++ memory management functionality. In parallel codes we align allocated
memory regions to cache lines (64-bytes) using the \texttt{posix\_memalign}
 alternative to \texttt{malloc}, to avoid false sharing.

\subsection{Benchmark Problems}

 We tested the algorithms on synthetic problems with fixed $n_{i}$
and $m_{i}$ and with random fixed orthonormal $F_{i}$ and $G_{i}$
and with $H_{i}=I$. The observations $o_{i}$ were also random. We
set $L_{i}=I$ and $K_{i}=I$. The use of orthonormal $F_{i}$ and
$G_{i}$ avoids growth or shrinkage of the state vectors and hence
overflows and underflows. We used two typical state-vector dimensions,
either $n_{i}=m_{i}=6$ or $n_{i}=m_{i}=48$. One particular test,
designed to clarify a specific hypothesis, used $n_{i}=m_{i}=500$.
We denote the common dimension of the state by $n$.

Running times of the parallel algorithms do not include the time to
build the array of steps, since in a parallel application, the array
would typically be constructed using parallel input-output mechanisms
(if the data is stored in files), using a parallel simulation code,
or using an outer nonlinear Gauss-Newton-type solver, as explained
in Section~\ref{subsec:Data-Structures}. In all cases, parallelizing
these computations is clearly outside the scope of this article.

\begin{figure*}
\hspace{0.1\columnwidth}\includegraphics[width=0.8\columnwidth]{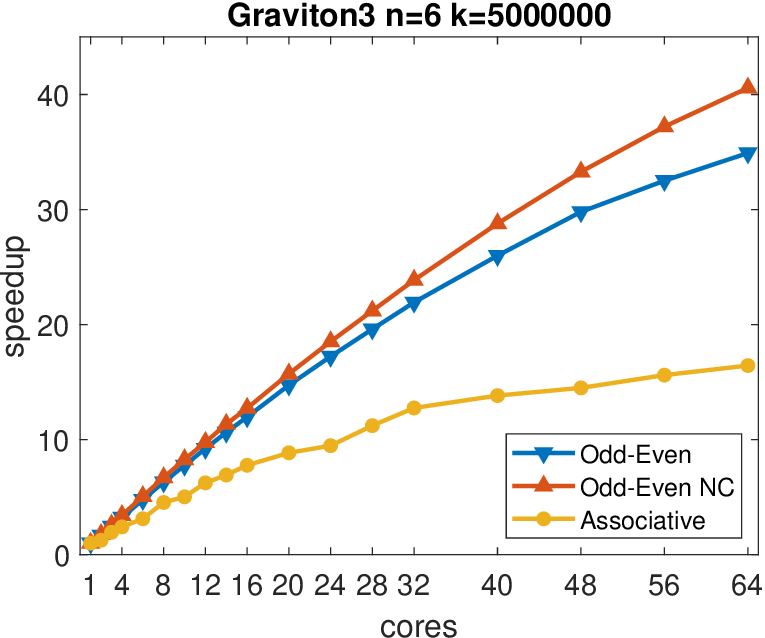}\hfill{}\includegraphics[width=0.8\columnwidth]{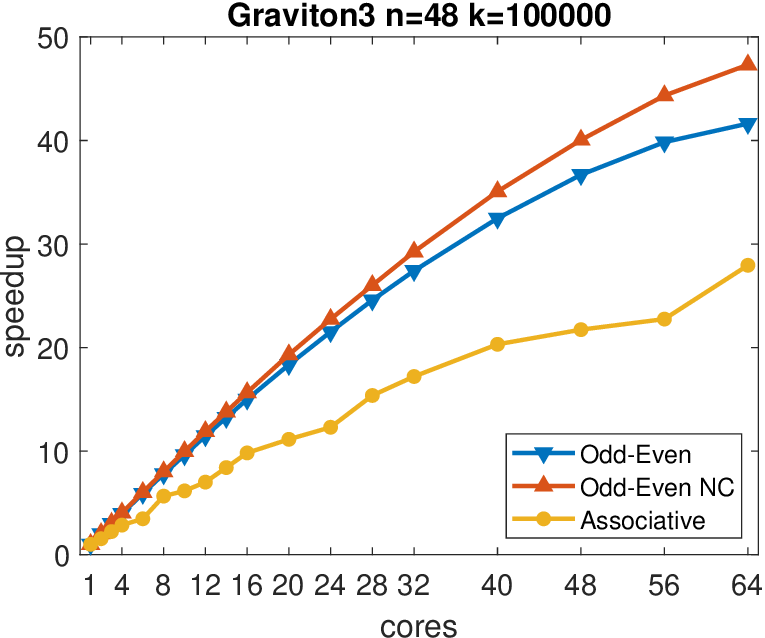}\hspace{0.1\columnwidth}\hphantom{}\\
~\\
\hphantom{}\hspace{0.1\columnwidth}\includegraphics[width=0.8\columnwidth]{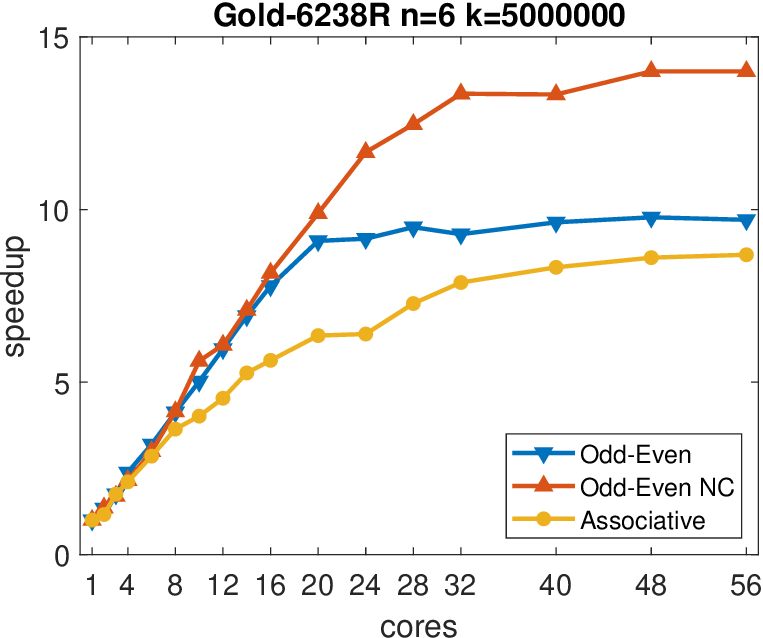}\hfill{}\includegraphics[width=0.8\columnwidth]{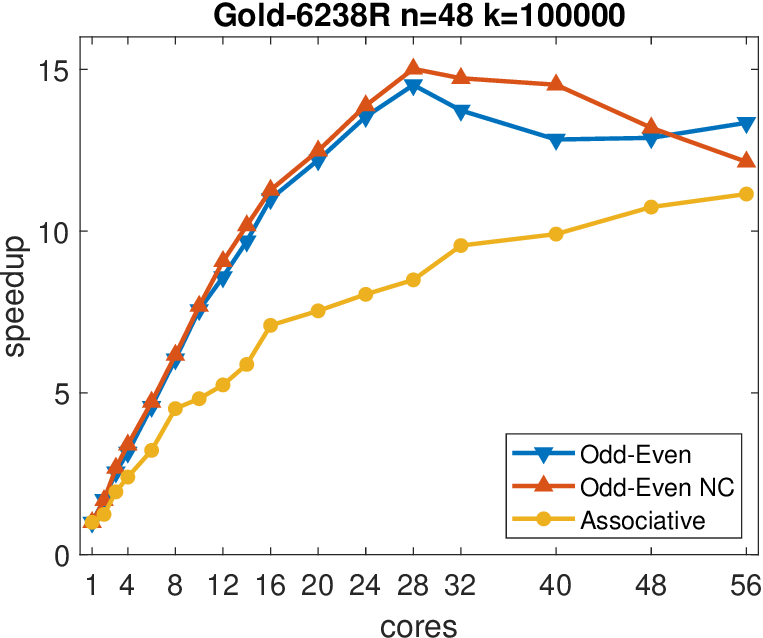}\hspace{0.1\columnwidth}

\caption{\label{fig:speedups}Speedups of the parallel smoothers . The ratios
are relative to the running time of the same implementation on 1 core.
The graphs are based on the same data shown in Figure~\ref{fig:running-times}.}
\end{figure*}

\subsection{Experimental Platforms }

We evaluated the algorithms on several multi-core shared-memory servers.
One has a Amazon Graviton3 CPU with 64 ARM cores running at 2.6~GHz
and 128~GB of RAM (AWS EC2 \texttt{c7g.metal} instance; this instance
runs directly on the hardware, with no virtualization). The second
has two Intel Xeon Gold 6238R CPUs running at 2.2~GHz and 200~GB
of DRAM. Each CPU has 28 physical cores for a total of 56.  We also
performed the experiments on a server with two older Intel Xeon CPUs,
model E5-2699v3 with 18 physical cores each running at 2.30GHz. 
The results are similar to those on the Intel Xeon Gold 6238R server
and are not shown.

We used GCC (version~7.5.0 on Intel servers and version 13.3 on ARM)
and Intel's threading building blocks, also known as TBB (version
2024.1 on Intel servers and 2021.11 on ARM). We used vendor-optimized
BLAS and LAPACK libraries, MKL version 2024.1 on Intel servers and
ARM Performance Libraries version 24.10 on the ARM server. Unless
otherwise noted, we used the single-threaded (and thread-safe) version
of the BLAS and LAPACK libraries.

\begin{figure*}
\begin{centering}
\hspace{0.1\columnwidth}\includegraphics[width=0.8\columnwidth]{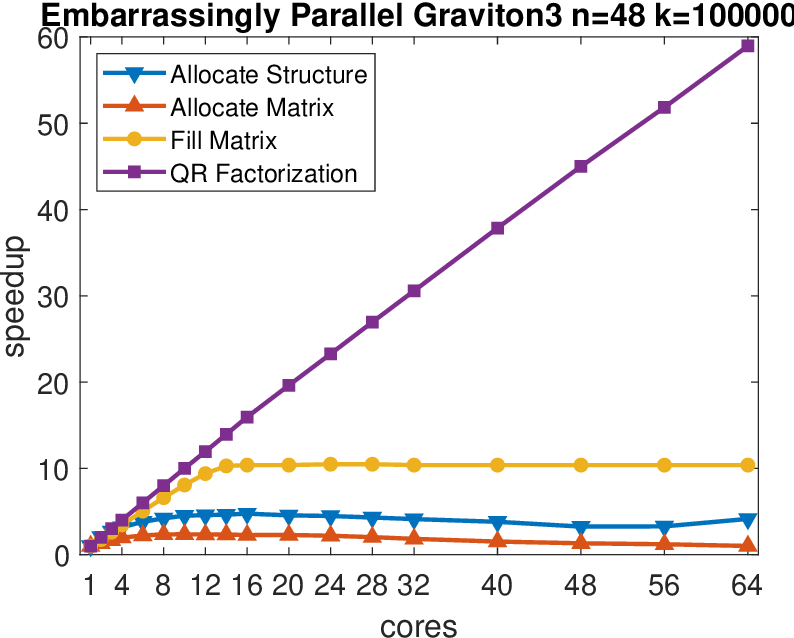}\hfill{}\includegraphics[width=0.8\columnwidth]{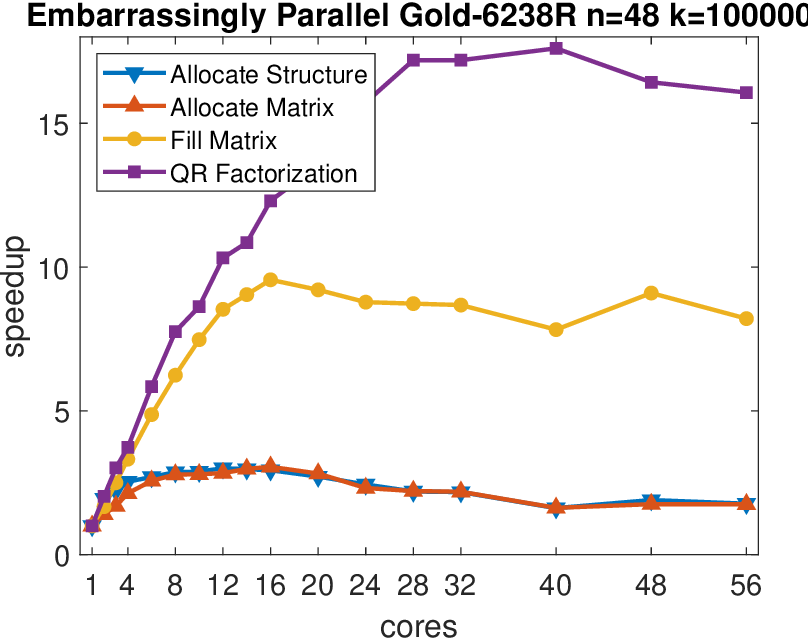}\hspace{0.1\columnwidth}
\par\end{centering}
\caption{\label{fig:ep-1}Speedups of 4 phases of a representative but embarrassingly-parallel
micro-benchmark.}
\end{figure*}

To characterize the capabilities of the hardware and of TBB, we implemented
a simple embarrassingly-parallel micro-benchmark whose 4 phases characterize
building blocks of our algorithms. In the first phase, the code allocates
$k$ structures that represent steps and stores their addresses into
an array. Next, for every step the algorithm allocates a $2n$-by-$n$
matrix and stores its address in the step structure. In the third
phase, the code fills every matrix $A$ with values $A_{ij}=i+j$.
Finally, the code computes the QR factorization of each matrix. Each
phase is implemented with a separate \texttt{tbb::parallel\_for}.
The block size is 8 to avoid false sharing in phase~1. The speedups,
shown in Figure~\ref{fig:ep-1} for $n=48$, indicate that speedups
on QR factorization at this size excellent on the ARM server (59x
on 64 cores; almost the same for $n=6$) but the memory-allocation
and memory-filling phases do not scale well. On the Intel server speedups
in the QR phase are limited to about 18 (and can be achieved with
a single CPU). On both servers, the memory allocation phases scale
poorly in spite of TBB's scalable allocator, but their running times
are fairly insignificant.

\subsection{Results}

The running times of the linear Kalman smoothers are shown in Figure~\ref{fig:running-times}
(all running times are medians of 5 runs) and the speedups of the
parallel smoothers are shown in Figure~\ref{fig:speedups}. Figure~\ref{fig:running-times}
shows the running time of our new parallel algorithm, denoted \emph{Odd-Even},
of the Särkkä \& García-Fernández algorithm, denoted \emph{Associative},
as well as of a conventional Kalman (RTS) smoother and of an implementation
of the Paige-Saunders sequential QR-based algorithm. The conventional
Kalman smoother and the parallel Associative smoother compute the
smoothed states and their covariance matrices together; they cannot
compute one without the other. In the Paige-Saunders algorithm and
in our Odd-Even parallel algorithm, the evaluation of the covariance
matrices of the smoothed states is a separate phase that we can skip,
and we did evaluate the performance without this phase. This is denoted
in the graphs by NC (no covariance). The NC variants are optimized
for use in Levenberg-Marquardt-based nonlinear Kalman smoothing~\cite{10.1109/ICASSP40776.2020.9054686}. 

We can draw several conclusions from this data. First, all the parallel
smoothers have a considerable overhead on one core. The parallel Odd-Even
algorithm is about 1.8--2.5 times slower than the sequential Paige-Saunders
algorithm (1.8--2.0 when covariance matrices are not computed). The
overhead of the Associative parallel algorithm relative to a conventional
Kalman smoother is similar, about 1.8--2.7.

Second, the Odd-Even parallel smoother is faster than the Associative
Kalman smoother (except sometimes on 1 core only). 

Third, the parallel algorithms do speed up and all of them easily
beat the sequential variants. The Odd-Even smoother exhibits better
speedups than the Associative one. On the ARM server performance improves
monotonically and appreciably with the number of cores. On the Intel
server, performance improves up to about 28 cores (one CPU) and mostly
stagnates beyond that.

\begin{figure*}
\hspace{0.1\columnwidth}\includegraphics[width=0.8\columnwidth]{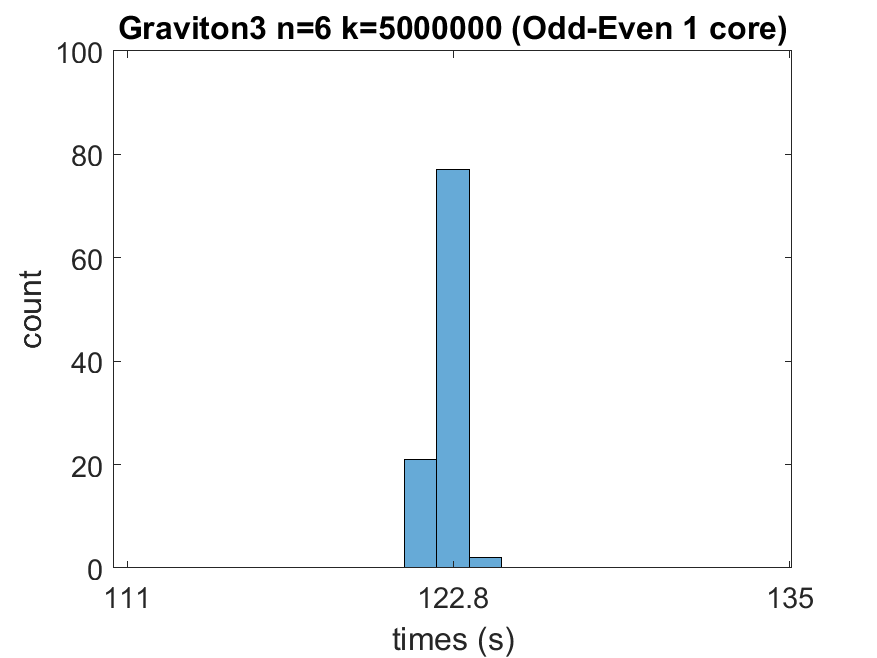}\hfill{}\includegraphics[width=0.8\columnwidth]{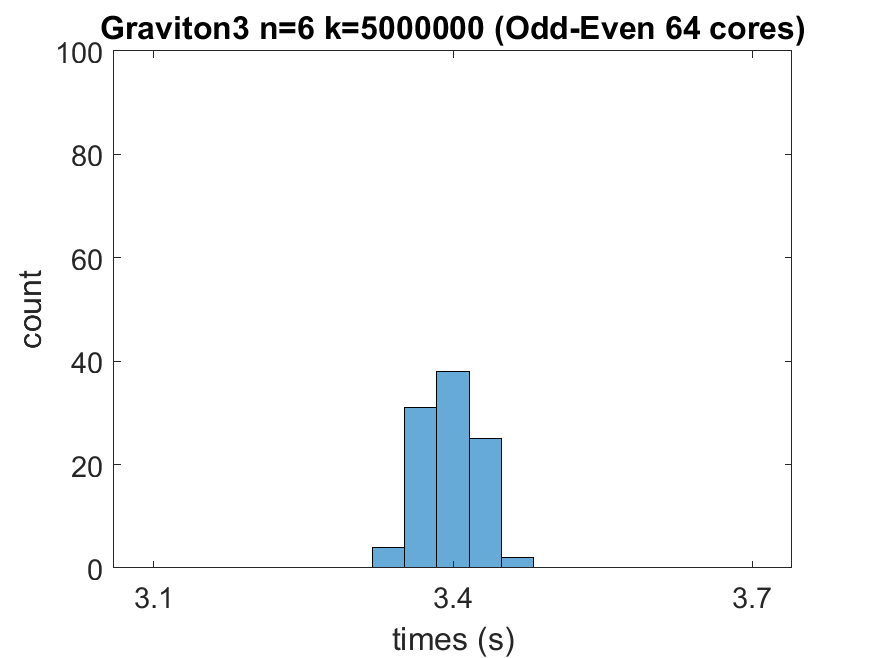}\hspace{0.1\columnwidth}

\caption{\label{fig:distribution}Running times distributions of the Odd-Even
algorithm on 1 core and on 28 cores. The histograms analyze 100 runs
each. The horizontal spans are set to 20\% of the median running time.}
\end{figure*}

Figure~\ref{fig:distribution} shows the relatively small effect
of the randomized work-stealing scheduler on the running time. Both
histograms display the distribution of 100 running times; the horizontal
span in both histograms is set to 20\% of the median running time.
With 64 cores, we observe running-time variations of up to $\pm$2.4\%
of the median running time. On 1 core the TBB scheduler was not invoked
at all and the maximum variation is smaller, less than 0.9\%. On the
Intel Xeon server, the variation with 28 cores is 13\% of the median
running time and with 1 core only 1.5\% (the graphs are omitted).

\begin{figure*}
\hspace{0.1\columnwidth}\includegraphics[width=0.8\columnwidth]{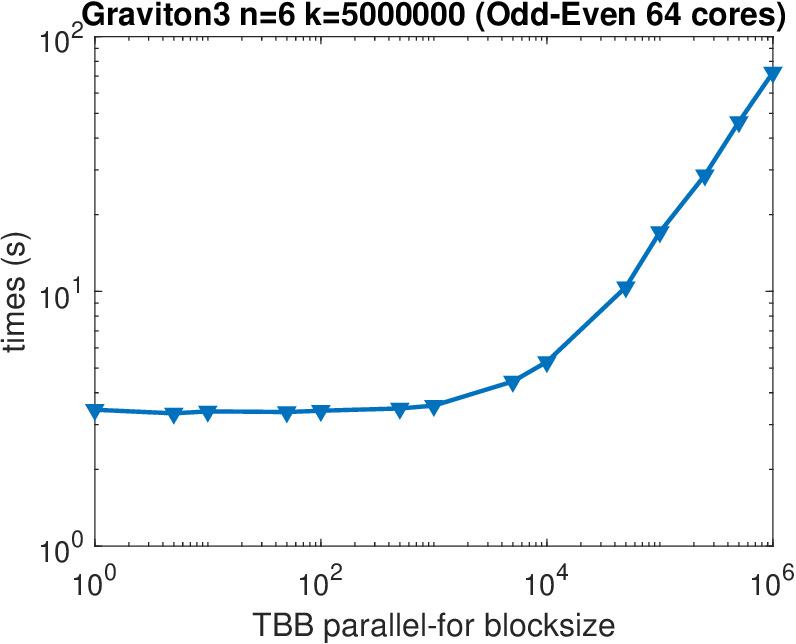}\hfill{}\includegraphics[width=0.75\columnwidth]{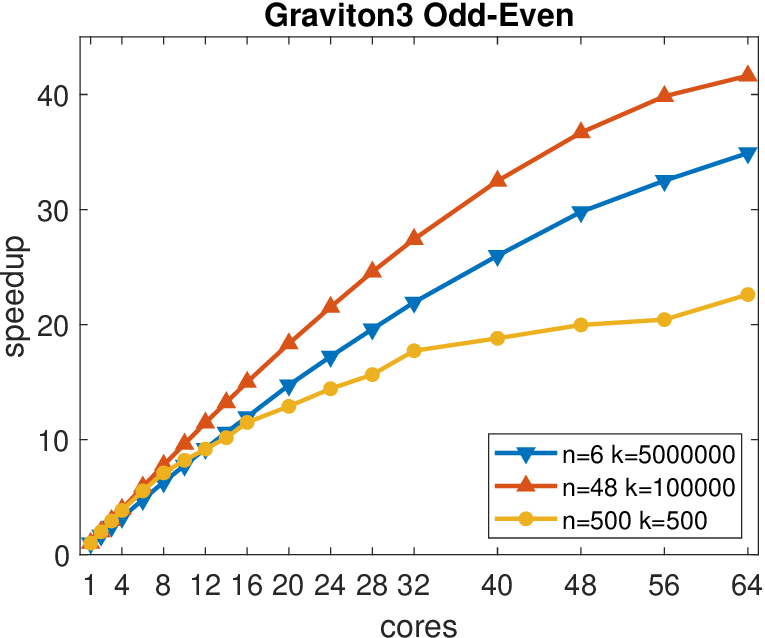}\hspace{0.1\columnwidth}

\caption{\label{fig:blocksize-problemsize}Left: Running times of the Odd-Even
algorithm on 64 cores as a function of the block-size parameter in
invocations of \texttt{tbb::parallel\_for} loops. Right: Speedups
of the of the Odd-Even algorithm problem of different dimensions,
with a TBB block size of 10 for $n=6$ and $n=48$ and with a block
size of 1 for $n=500$.}
\end{figure*}

The graph in Figure~\ref{fig:blocksize-problemsize} (left) shows
the effect of the block-size parameter in invocations of \texttt{tbb::parallel\_for}
loops. The graph shows that even for small dimensions ($n_{i}=6$),
performance remains roughly same with block sizes ranging from 1,000
all the way down to 1. At block sizes between 5,000 and 1,000,000,
the smoother does slow down due to insufficient parallelism, as expected.

The graph in  Figure~\ref{fig:blocksize-problemsize} (right) shows
what limits scalability on different problem sizes. The speedups on
a problem with $n=48$ are somewhat better than with $n=6$. This
is most likely caused by the better computation-to-communication ratio
at $n=48$. On an even larger state dimension $n=500$ but a smaller
number of steps $k=500$, the speedups are worse than the other two
cases, most likely due to insufficient parallelism.

\section{\label{sec:Conclusions}Conclusions}

The parallel odd-even linear Kalman smoother that we introduced in
this article offers best-in-class performance along with functionality
not available until now in a parallel algorithm. Given enough cores
our new parallel algorithm outperforms all sequential Kalman smoothers,
in spite of its additional arithmetic overhead. The algorithm also
outperforms the parallel-in-time Kalman smoother proposed by Särkkä
and García-Fernández~\cite{10.1109/TAC.2020.2976316}.

The core of the new algorithm is a specialized QR factorization. The
use of this framework, introduced by Paige and Saunders~\cite{PaigeSaunders:1977:Kalman},
contributes to the functionality of the smoother in several ways.

First, the algorithm is conditionally backward stable. The backward
stability depends (only) on the input covariance matrices, the $K_{i}$
and $L_{i}$ matrices, just like the stability of the Paige-Saunders
algorithm. In particular, when these matrices are either diagonal,
a common case, or well conditioned, the overall algorithm is backward
stable. In contrast, nothing is known about the numerical stability
of the only other parallel-in-time Kalman smoother~\cite{10.1109/TAC.2020.2976316}. 

Second, the algorithm can handle problems in which the expectation
of the initial state is not known. This is a fairly common case that
arises, for example, in inertial navigation. Our algorithm can also
handle cases where $H_{i}\neq I$. The Särkkä and García-Fernández
smoother~\cite{10.1109/TAC.2020.2976316} cannot handle such problems,
but unlike ours, it can handle problems with singular input covariance
matrices, just like the conventional Kalman (RTS) smoother.

Third, the covariance matrices of the smoothed state estimates are
computed in a distinct final phase. Our implementation can skip this
phase, speeding up the computation when these covariance matrices
are not needed, which is the case in a Levenberg-Marquart nonlinear
Kalman smoother~\cite{10.1109/ICASSP40776.2020.9054686}. The Särkkä
and García-Fernández smoother must compute the covariance matrices,
just like conventional Kalman smoothers, so it cannot benefit from
this optimization.

Another key contribution of this article is the discovery that the
SelInv algorithm~\cite{10.1145/1916461.1916464} can be used to compute
the covariance matrices of the smoothed estimates. This discovery
also applies to the Paige-Saunders algorithm, where SelInv can replace
a sequence of orthogonal transformations, but it is particularly crucial
for the odd-even parallel algorithm, because there is no apparent
way to extend the Paige-Saunders approach to this case.

The final contribution of this paper is the parallel implementation
of the Särkkä and García-Fernández and the first detailed report
on its scaling. 

We do acknowledge one limitation of the algorithms (both ours and
that of Särkkä and García-Fernández): due to the arithmetic (work)
overheads of the parallel algorithms, the sequential variants are
faster on small number of cores. Since parallelizing the operations
of each step in a sequential-in-time algorithm appears to be effective
only in very high dimensions~\cite{10.1109/TCST.2011.2176735}, developing
low-overhead parallel-in-time smoothers appears to be an important
open challenge. 

We also acknowledge that both families of algorithms exhibit limited
strong scaling; the data in Figure~\ref{fig:ep-1} suggests that
this is mostly due to memory-bandwidth and cache misses, but also
that it might be possible to further improve performance. 

Finally, it is evident from the structure of $UA$ that $(UA)^{T}(UA)$,
the coefficient matrix of the normal equations of the linear least-squares
problem, is block tridiagonal. Therefore, the normal equations can
be solved in parallel using block odd-even reduction of this block
tridiagonal matrix~\cite{10.1137/0707049,10.1137/0713042}, yielding
a third parallel algorithm for Kalman smoothing. However, this approach
is unstable and does not appear to have any advantage over our new
algorithm.

Our implementations are available at\emph{ \url{https://github.com/sivantoledo/ultimate-kalman}}
under standard open-source licenses.

\noindent \textbf{Acknowledgments}. We thank the reviewers for comments
and suggestions that helped improve the article. This research was
supported in part by grant 1919/19 from the Israel Science Foundation.

\clearpage

\noindent 

\bibliographystyle{IEEEtran}
\bibliography{kalman}

\end{document}